# An efficient Search Tool for an Anti-Money Laundering Application of an Multi-national Bank's Dataset

Nhien-An Le Khac[1], Sammer Markos[1], Michael O'Neill[1], Anthony Brabazon[1] and M-Tahar Kechadi[1]
[1]School of Computer Science & Informatics, University College Dublin, Dublin, Ireland

**Abstract -** Today, money laundering (ML) poses a serious threat not only to financial institutions but also to the nations. This criminal activity is becoming more and more sophisticated and seems to have moved from the cliché of drug trafficking to financing terrorism and surely not forgetting personal gain. Most of the financial institutions internationally have been implementing anti-money laundering solutions (AML) to fight fraud activities. However, the AML systems are so complicated that simple query tools provided by current DBMS may produce incorrect and ambiguous results and they are also very time-consuming due to the complexity of the database system architecture.

In this paper, we present a new approach for identifying customers quickly and easily as part of an AML application. This will help AML experts to identify quickly customers who are managed independently across separate databases of the organization. This approach is tested on large and real-world financial datasets. Some preliminary experimental results show that this new approach is efficient and effective.

## 1 Introduction

Money laundering (ML) is a process of disguising the illicit origin of "dirty" money and makes them appear legitimate. It has been defined by Genzman as an activity that *"knowingly engage in a financial transaction with the proceeds of some unlawful activity with the intent of promoting or carrying on that unlawful activity or to conceal or disguise the nature location, source, ownership, or control of these proceeds"* [7]. Through money laundering, criminals try to convert monetary proceeds derived from illicit activities into "clean" funds using a legal medium such as large investment or pension funds hosted in retail or investment banks. This type of criminal activity is getting more and more sophisticated and seems to have moved from the cliché of drug trafficking to financing terrorism and surely not forgetting personal gain. Today, ML is the third largest "Business" in the world after Currency Exchange and Auto Industry. According to the United Nations Office on Drug and Crime, worldwide value of laundered money in a year ranges from $500 billion to $1 trillion [1] and from this approximately $400-450 Billion is associated with drug trafficking. These figures are at times modest and are partially fabricated using statistical models, as no one exactly knows the true value of money laundering, one can only forecast according to the fraud that has already been exposed. Nowadays, it poses a serious threat not only to financial institutions but also to the nations. Some risks faced by financial institutions can be listed as reputation risk, operational risk, concentration risk and legal risk. At the society level, ML could provide the fuel for drug dealers, terrorists, arms dealers and other criminals to operate and expand their criminal enterprises. Hence, the governments, financial regulators require financial institutions to implement processes and procedures to prevent/detect money laundering as well as the financing of terrorism and other illicit activities that money launderers are involved in. Therefore, anti-money laundering (AML) is of critical significance to national financial stability and international security..

Typically, an AML system is composed of some components such as customer identification, transaction monitoring, case management, reporting system, etc. Among them, customer identification is one of the most important tasks as it assists AML experts in monitoring customer behaviours; transactions that they are involved in, their frequencies, values, etc. Fundamentally, a customer is identified by searching customer databases using query tools provided by DBMS. However, in the case where a specific customer is stored in separate databases that are managed independently, this will require a very large processing time due the search operations initiated over all the databases. Users need firstly to login to different databases, run the same query repeatedly, get the results separately, and displayed independently. Furthermore, in large financial institutions, these databases are heterogeneous and have very complex designs. This sort of approach allows great flexibility, however it has poor performance. In addition, data quality is also another factor that makes this naïve approach becoming unfeasible.

In this paper, we present a new approach for identifying customers in an international investment bank ***BEP***[1]. This approach provides a global view of customer information and it is developed as a tool that allows the users to quickly and efficiently identify customers who are managed independently across separate databases. This tool is a component of an AML solution developed for ***BEP***.

---

[1] Real name of the bank can not be disclosed because of confidential agreement of the project.

The rest of this paper is organised as follows: the section 2 presents a background highlighting the current status of ***BEP***'s datasets and their customer search problems within the AML context. Some indexing approaches are also discussed in this section. We present our new approach that is a global indexing based on word-ordered grouping and inverted list in the Section 3. We describe the implementation of this approach in Section 4. Section 5 presents preliminary experimental results. Finally, we conclude in section 6.

## 2 Backgrounds

We start this section with a brief presentation of an AML project at ***BEP*** and then we will discuss on customer search problems in its current environment. We finish this section by reviewing some indexing approaches for data search in the literature.

### 2.1 AML in BEP

Similar to any banking institution, ***BEP*** is required by law to conduct strict AML governance on all transactions. The ***BEP*** AML Unit does not have an automated solution to support pattern recognition and detection of suspicious activities. The purpose of this project is to apply new principles and methodologies to build an AML framework in order to detect suspicious customer transactions and behaviour for the AML Unit. In this framework, one of the important components is customer identification. Before launching any customer transactional investigation, the customer should be identified in all customer databases of ***BEP***. The structure of the ***BEP*** databases is complicated and there are many problems with data quality that will can be extracted and analysed, which are discussed in the following paragraphs.

***BEP*** datasets are divided into different environments corresponding to sixteen clients with multiple funds per client and managed hence by sixteen independent databases. When a new customer or an investor X want to invest into a specific fund (client specific), the AML team would request certain documentation and will always treat him as a new customer even though he could already invest into one/more of the other fifteen clients, i.e. already exist in another databases. The purpose of the customer search is to verify and identify a customer's profile in all invested funds. The AML Unit is currently applying a manual search based on DBMS queries. However, this is a time-consuming task because users should login separately to each database and carry out repeated queries. Moreover, each database contains not only data but also its meta-data, so many joint operations are needed to retrieve the information required.

Meanwhile, the data quality is also another impact that affects the searching task. ***BEP***'s input GUI is not efficient and its databases design is cumbersome. Each customer database is "identical", i.e. the customer identification (CID) is only unique in this database but the CID is not unique in all databases. For instance, we can have (name= "John Smith", CID= ***"12345"***) in database A vs. (name= "Peter Chang", CID= ***"12345"***) in database B. Briefly; there is a uniqueness violation at the global level. Furthermore, each database has a different set of quality problems at the instance level. Some problems can be listed as:

- Missing values, dummy values or null. These appear in most of the data fields in all databases except the CID, the customer type (corporate, individual and joint) and the fund name.

- Misspellings; usually typos and phonetic errors. For instance, we have "MACAO" vs. "MACAU", "11 1101" vs. "11-1101", "Bloggs Corporation A/C 001" vs. "Bloggs Corporation 001", etc.

- Abbreviations; e.g. "A/C" vs. "AC" and "Account"

- Word transpositions; e.g. "John Smith" vs. "Smith John"

- Duplicated records e.g. "John Smith" vs. "J. Smith"

Moreover, the names of some corporate customers are normally not identical even though they are the same company. For instance, "First Commercial Bank Ltd"[2], "First Commercial Bank Ltd OBB Account", "First Commercial Bank Ltd Trust Account TA 101010", "First Commercial Bank Ltd Trust Account TA 505055", etc. We call this a "*company name group*" property. Besides, some customer databases also have the problem of *incoherent data* in address data fields. The address information includes the following data fields: "Street", "Town", "Zip", "Country Code" and "Country". And then, for example, the "Zip" field contains information about the street, house number and/or town, city instead of its zip code.

Because of the customer datasets quality as well as its complicated design, the manual customer search task by DBMS queries currently takes more than two hours to identify a customer.

### 2.2 Indexing

Fundamentally, search engines index the data in order to facilitate fast and accurate information retrieval. Some indexing methods in literature are tree-based, suffix tree,

---

[2] Again, due to the confidential agreement, all examples presented in this paper do not use the real customer names, company names and address

inverted list, citation index, ngram index, term document matrix, etc. The tree-based index would be the most popular method where the search operations are linked with tree nodes. The tree topology can be varied from a binary [10] to a B-Tree family such as B/B*/ B+Tree [2] [3] [12]. For instance, some DBMS implement an index structure based on B-Tree such as MySQL, SQL Server [11]. Nevertheless, this topology is not efficient enough for indexing complex, heterogeneous, and bad quality data fields.

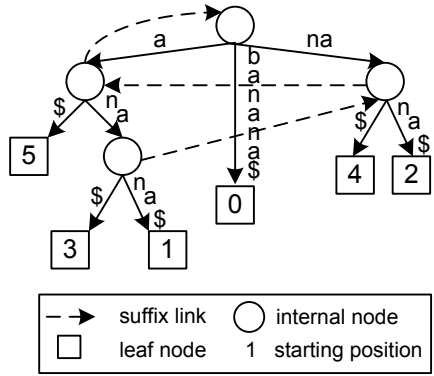

**Figure 1. A suffix tree for "banana$"**

Suffix tree [8], so-called PAT tree or position tree, is a data structure that presents a given string in a suffix way (Figure 1). The suffix tree for a string S is a tree whose edges are labelled with strings such that each suffix of S corresponds exactly to one path from the tree's root to a leaf. The advantage of suffix tree is that operations on S and its substring can be performed quickly. However, the constructing suffix tree takes time and storage space linear in the length of S.

Inverted list [13] is another kind of index where each entry in the index table includes two elements: an atomic search item and a list of occurrences of this item in the whole search space. For example, the index of a book lists every page on which certain important words appear. This approach is normally implemented by the hash table [4] or the binary tree [10]. Inverted list is one of the most efficient index structures [14].

Citation index approach [6] stores the citation or hyperlinks between documents to support citation analysis. This approach is normally applied in the Bibliography domain. Ngram index [9] stores sequences of length of data and term document matrix stores the occurrences of words in the documents in a two-dimensional sparse matrix. The last two index methods are mainly used in information retrieval or text mining [5].

## 3 Customer Search approaches

As mentioned above, the current techniques based DBMS queries are not suitable for the ***BEP***'s AML system, as they depend strongly on the quality of the data sets. Therefore, the current quality of ***BEP***'s customer data sets should be improved before running any query. For instance, in order to correct the misspelling problem, a spelling module should be implemented to deal with typos and phonetic errors. Similarly, abbreviated words must be uniform across all separate databases; e.g. "A.C", "Account", "AC." are transformed to "A/C". Meanwhile, data mining techniques such as decision tree induction, regression, and inference-based tools can be applied to fill missing values (tuples that contain missing value fields cannot be ignored because all customer information are important). Indeed, in some cases, we should fill the missing value manually. Similarity, the word transposition and duplicated records often need manual intervention. Besides, the incoherent data problem in address data fields (ref. Section 2.1) can only be manually corrected but it is an unfeasible task with large datasets. Briefly, a general solution for improving efficiently the quality of ***BEP's*** customer data sets is still an open question. Finally yet importantly, the execution of DBMS queries on sixteen independent ***BEP's*** customer databases is also a very time-consuming task. Next, we present our approach, which can overcome the quality and design problems of ***BEP's*** customer databases

### 3.1 Basic concepts

In this new approach, we aim to provide a global view of information about all customers managed independently across the ***BEP's*** customer databases. Concretely, we build a global index of these customers and provide a search engine for AML users. Firstly, by analysing ***BEP's*** customer datasets, some important features can be summarised as:

- There are two main types of customer: individual and corporate. The individual customer has two name fields: "First Name" and "Last Name". In some records, "First Name" (resp. "Last Name") field stores all parts of customer name; e.g. in a record X, "First Name" field stores "John Smith" and its "Last Name" is empty. This is a special kind of missing value. The corporate customer has only one name field: "Company Name" and most of them have a "company name group" property as mentioned in the Section 2.1 above.

- The "Country" field is the most popular, i.e. its missing value is less than 1%.

In addition, we also build a summary of all abbreviation cases after this pre-processing process. We build our solution based on these features and assuming that all abbreviation words are expressed in a uniform way as well as all minor data preparation is performed.

In order to deal with the incoherence of the address data (Section 2.1), we merge all of the address data fields into one field called "*Address*". Similarity, we merge the "*First*

Name" and "Last Name" of individual customer into one field named *"Customer Name"* to solve the missing value problem. For each data record, word order in *"Address"* and *"Customer Name"* can vary. For instance, we can have "John Smith" vs. "Smith John" or "123, Main Street" vs. "Main Street, 123" (word transposition problem). Therefore, the inverted list technique can be used in this case to index *"Address"* and *"Customer Name"*.

Meanwhile, the word order in the *"Company Name"* field is important because of the *"company name group"* property. Therefore, we need another type of index in this case. We can address the *"company name group"* as a kind of suffix problem and we can then use the suffix-tree topology. Nevertheless, the implementation of this topology is complicated. Hence, we rely on this topology to build an index tree, which is simpler than the suffix tree for the *"Company Name"* field of **BEP**'s customer datasets. So, our global index is composed of three main parts: *"Company Name"* index, *"Customer Name"* and *"Address"* index and they are based on tree topology (the first index) and inverted list (the last two ones), which are detailed in the following section.

## 3.2 Index architecture

The main architecture of our index consists of "Company Name" tree, "Customer Name" and "Address" inverted lists (Figure 2, 3 and 4). Furthermore, the whole customer datasets are grouped by the customer type (corporate or individual) and by the country.

**Company Name Tree (CN tree) design.** This is a suffix tree based topology. Generally, the first word of a company name appears at the root level (level 0) and its last word is at the leaf level (Figure 2). The *CN tree* includes a set of nodes. Each node contains one or many elements and each element at level $l$ links to only one node at level $l+1$ or to NULL if $l$ is at the leaf level. Each element has a key, which is a word from the "company name" string. Hence, each node $p$ at the level $l$ ($l>0$) contains all words derived from their prefix word at the level $l-1$ and so on. Formally, supposing that a customer datasets includes a set of $n$ company names $N$ and each company name $cn_i \in N$ (i=1, 2 ... n) is a string composed of a set of words $w_j$ and the number of words in a $cn_i$ is noted by $C_i$, a *CN tree* $T$ of $N$ from customer datasets is defined by the following:

- The height of $T$ is $h$, $h = max(C_i)$, $i = 1,2...n$

- $\forall$ node $p_k \in T$, $p_k \supseteq$ set of elements $\{em\}$: Card$\{em\}>0$, $em \equiv w_j$.

- The level $0$ has at most one node $p0$.

- Each element $em0_i \in p0$, $em0_i$ contains the first word of each company name $cn_i$.

- $\forall$ element $em \in T$ at the level $l$ ($l > 0$), there is a link, so-called node-link, between $em$ and a node $p_{em} \in T$ at the level $l+1$. The node $p_{em}$ contains the first word of all suffixes of the word $w_j$ stored in $em$.

- A path from the root to the leaf by following node-links will create a specific company name. Each element at the leaf level does not have a node-link but a list of {client identification (FID), customer identification (CID)} of the company name that creates this path.

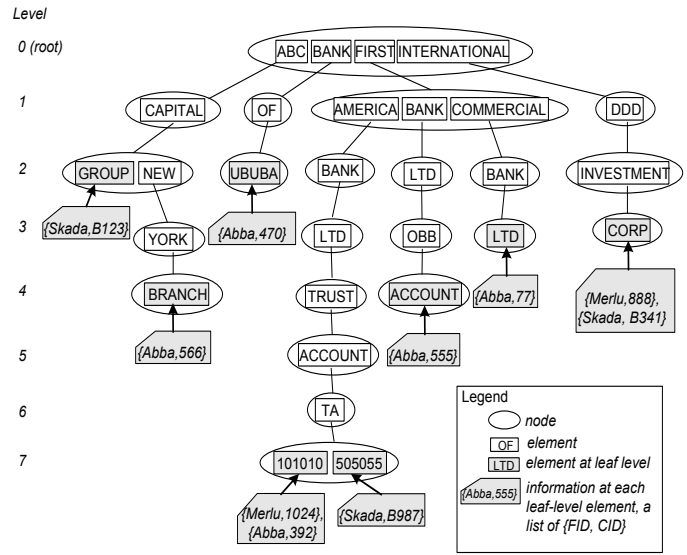

**Figure 2. An example of CN tree index**

For instance, as shown in Figure 2, we have the following company name: "FIRST COMMERCIAL BANK LTD", "FIRST BANK LTD OBB ACCOUNT", "FIRST AMERICA BANK LTD TRUST ACCOUNT TA 101010", "FIRST AMERICA BANK LTD TRUST ACCOUNT TA 505055", "ABC CAPITAL GROUP", "ABC CAPITAL NEW YORK BRANCH", "BANK OF UBUBA" and "INTERNATIONAL DDD INVEST CORP". Hence, the root node has four elements: $em0_0$ = "ABC", $em0_1$ = "BANK", $em0_2$ = "FIRST" and $em0_3$ = "INTERNATIONAL". The element $em0_0$ links to a node that has one element "CAPITAL" at the level 1. Then, this element links to another node that has two elements: "GROUP" and "NEW". The element "GROUP" is at the leaf level, it then contains {"Skada", "B123"} (list of {FID, CID} has only one element). Otherwise, the element "NEW" links to another node at the level 2, etc. The path from the root node with the element "ABC" following its node-links creates two company names "ABC CAPITAL GROUP" and "ABC CAPITAL NEW YORK BRANCH". Similarity, the element $em0_2$ links to a node with three elements at the level 1: "AMERICA", "BANK" and "COMMECIAL". The element "AMERICA" links to another node at the level 2 and so on, at the level 7, the element "101010" is at the leaf level and contains [{Merlu, 1024}, {Abba, 392}] (list of {FID, CID} has two elements).

Based on this *CN Tree*, the search engine can find all {FID, CID} of a requested "*company name*". For instance, if the query is "ABC CAPITAL GROUP" then the result is {"Skada", "B123"}, etc. Indeed, this index also supports an approximate search i.e. users might not know the sufficient name of a company so they just input its first few worlds, e.g. the query is "ABC CAPITAL" then the result list will be {"Skada", "B123"} and {"Abba", "566"}. Next, we can retrieve all details of customers whose {"FID", "CID"} are in the result list.

| Items (words) | List of {Fund ID, Customer ID} |
|---|---|
| …… | …… |
| John | {"Abba", "1234"}, {"Merlu", "112"} |
| …… | …… |
| Murphy | {"Merlu", "112"} |
| …… | …… |
| Smith | {"Abba", "1234"} |
| …… | …… |

**Figure 3. An example of Customer Name Index**

"**Customer Name Index**" is an index table based on the inverted list technique. This index table consists of two parts: items and a collection of lists; one list per item (Figure 3). An item is a word from the "*Customer Name*" i.e. each customer name in customer datasets is parsed into a set of separate words. For instance, the customer name "John Smith" is split into two words: "John" and "Smith". A list $L_i$ of a word $w_i$ records tupes of {FID, CID} of customers whose names contain the word $w_i$. We have, for example, the customer "John Smith" with {FID= "ABBA", CID= "1234"} and "Murphy John" with {FID= "MERLU", CID= "112"}. Hence, the index table has three elements: ["John": {"ABBA", "1234"}, {"MERLU", "112"}], ["Murphy": {"MERLU", "112"}] and ["Smith": {"ABBA", "1234"}].

"**Address Index**" is also an index table based on inverted list technique. Similar to the "*Customer Name Index*", its index table consists of two parts: items and a collection of lists, one list per item (Figure 4). An item is a word from the "*Address*" i.e. each address in customer datasets is also parsed into a set of separate words. For instance, we have an "*address index*" table as shown in Figure 4.

The whole index structure is shown in Figure 5. In order to limit the search space, these datasets are also grouped by country (the most popular data field, ref. section 3.1). Therefore, our search engine allows users to launch requests on customer information managed in different databases only through their name and address. If a customer is a corporate then the search process will scan the *CN Tree* and *Address Index table*. Meanwhile, *Customer Index Table* and *Address Table* are used for an individual customer. Advantages and problems of this approach will be discussed in section 5.

| Items (words) | List of {Fund ID, Customer ID} |
|---|---|
| …… | …… |
| 123 | {"Abba", "1234"}, {"Skada", "347"} |
| …… | …… |
| Avenue | {"Merlu", "112"} |
| …… | …… |
| Sunset | {"Abba", "1234"} |
| …… | …… |

**Figure 4. An example of Address Index**

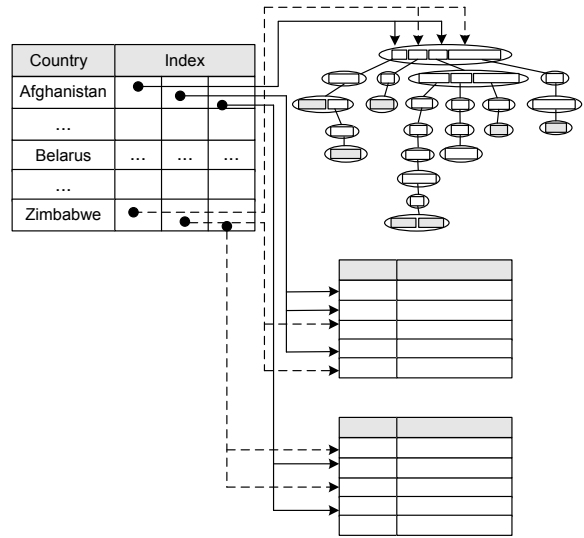

**Figure 5. Index structure**

## 4 Implementation

We developed our approach as a search tool based on a distributed paradigm and this tool is implemented as web services that can support 2-tier or 3-tier application model (Figure 6). We implemented services for two kinds of users: end-users and administrative users. There are indexing, updating services for administrators. End-users exploit this system through searching and extracting services.

### 4.1 Indexing service

This service scans all the BEP's databases once and builds indexes of "*Company Name*", "*Customer Name*" and "*Address*". Elements in each node of "*Company Name*" index as well as items in "*Customer Name*" and "*Address*" index table are sorted by lexical order. Indexing service also builds a country list and each element in this list stores information (hash code) about appropriate entries of three indexes above. These indexes are organised is main memory and this service allows them to be saved in secondary memory. Hence, the administrator only needs to create indexes once and stores them in databases (index databases) and then each time she/he reloads it to the main memory on application launch. In the real world of banking application, these indexes are loaded permanently in the main memory of the servers and are synchronised periodically with their databases (index databases). The customer information is not real-time data processing i.e. when a customer opens an account s/he always has to wait for a certain period of time for all the security checks to be carried out (7 days, for example) before the account is activated to perform her/his first transaction (or Subscription in term of the investment banking). Therefore, if indexes in the main memory are damaged due to system halt, the electric cut, etc., the administrators can reload them from index databases without losing any information.

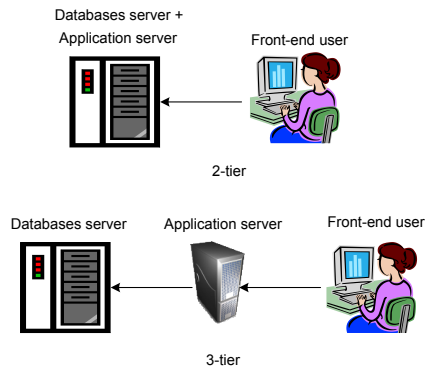

**Figure 6. Application models**

## 4.2 Updating service

When indexes are being exploited, new customer profiles are added in the customer datasets. Therefore, this service allows updating new customer information into indexes. It updates firstly in the main memory of the servers and it then synchronises this information with the index databases. The update service is automatically performed at a predefined time by scanning all the databases (all update information of customer is always stored for auditing).

## 4.3 Searching and extracting service

A search request submitted by the users includes customer/company name and its address. Searching service uses this information to look for a set of related {CID, FID} on "Company Name"/ "Address" indexes (corporate customer) or on "Customer Name"/ "Address" (individual customer). This is followed by performing queries on databases to retrieve related customer information. The users can choose which information to extract or perform further investigations.

## 5 Evaluation and discussion

We implemented and tested our approach on real-world customer datasets. The database architecture is similar to *BEP*'s databases. The hardware platform for testing includes 1 Pentium Dual Core 3.4Ghz 2Gb RAM Windows Server 2003 (database server), 1 Pentium 4 Hyper Threading 3.4Ghz 1Gb Windows XP SP2 (application server), 1 Pentium 4 2.7Ghz Windows XP SP2 512Mb (front-end user). This web service-based tool is developed in C#/Visual Studio 2005 and we use SQL Server 2005. All services are implemented at the application server and the database server manages the datasets (3-tier model, Figure 6). The number of records is approximately 32000 for all the databases.

We ran different tests on this platform and took the average results. The indexing time *I* is about *17* seconds. The total search time *S* is about *15* seconds (*15*s *40*ms) for one request. The search time *S* is composed of local search on the indexes, query process by SQL server and communication overheads; among them, the local search on the indexes only takes about *2* milliseconds on the application server. We also launched a customer search by SQL queries with exact "customer/company name" and "address" automatically on all the databases and it takes about 3 minutes for one request. We can see that our technique is much more efficient.

The approach presented in this paper has many advantages. First, it solves the problem of access and querying independent and separate databases by providing a global view of all customer information without changing the current architecture of *BEP*'s databases system. Then, this approach also overcomes the data quality problems that normally take an important time to pre-process, especially the manual correction of incoherent address data. The preliminary tests show that it is efficient. It is about 10 times faster than the normal approach by DBMS queries and exhibits better accuracy than the traditional approach.

Moreover, our approach also supports parallel processing where two threads can be launched to search independently on "Company/Customer Name" index and "Address" index. It can benefit from the multi-core architecture of BEP's servers.

Besides, two main aspects of this approach need to be improved. The first one is memory consumption because all index structures are stored in the main memory. However, each item stored is not a word but its hash code and it uses a small amount of memory in our experiments. Furthermore, in the real BEP's servers, the main memory space is greater than 100 Gb and the whole indexes take less than 0.2%. Besides, loading index in the main memory is normal practice of many current DBMSs to exploit it efficiently. Another aspect is the replication of items in the "Company Name" tree, e.g. the word "BANK" exists in many nodes (Figure 2). This problem can be improved by replacing this tree with a graph where each item appears once as a node and the edges linking these items represent the paths.

## 6 Conclusions and Future Works

In this paper, we have presented an approach for identifying specific customer patterns in an investment bank. This approach has been developed as a tool, which is a set of web services on a distributed platform. The contribution of this research is to provide a set of indexes combined with suffix-tree based on an inverted list in order to overcome the problem of database design and data quality of *BEP*'s customer datasets. Our experimental results obtained on parts of the *BEP*'s customer datasets, we can conclude that approach is very efficient tool and it satisfies the needs of an AML task.

Experimental results on real-platforms of BEP are also being produced and these will allow us to test and evaluate the tool robustness. We are currently working on the graph index approach that takes in account the memory consumption issue to tackle huge datasets. A multithreading version is also under development.